\documentclass[10pt,twocolumn]{article}

\usepackage[margin={1.8cm,2cm}]{geometry}
\usepackage{booktabs}
\usepackage{mathtools}
\usepackage{times}
\usepackage{framed}
\usepackage{authblk}
\usepackage[colorlinks,hidelinks]{hyperref}

\newcommand{\xhdr}[1]{\vspace{2mm}\noindent{{\bf #1}}}

\usepackage[numbers,square,sort]{natbib}
\usepackage{adjustbox}
\usepackage{array}

\DeclareMathOperator*{\argmax}{\mathrm{argmax}}

\usepackage{enumitem}
\setlist[description]{labelindent=0mm,leftmargin=0\parindent,listparindent=\parindent,parsep=\parskip}
\setlist[itemize]{leftmargin=\parindent,listparindent=\parindent,parsep=0.5\parskip}
\setlist[enumerate]{leftmargin=\parindent,listparindent=\parindent,parsep=0.5\parskip}

\begin{document}

\title{Addressing Complex and Subjective Product-Related Queries with Customer Reviews}
\newcommand{\modelname}{\emph{Moqa}}
\newcommand{\modelnamelong}{Mixtures of Opinions for Question Answering}
\newcommand{\eq}[1]{(eq.~\ref{#1})}

\author[1]{Julian McAuley\thanks{jmcauley@ucsd.edu} }
\author[2]{Alex Yang\thanks{alexyang@fb.com}}
\affil{Department of Computer Science, UC San Diego}

\date{\today}

\maketitle
\begin{abstract}
Online reviews are often our first port of call when considering products and purchases online. When evaluating a potential purchase, we may have a specific query in mind, e.g. `will this baby seat fit in the overhead compartment of a 747?' or `will I like this album if I liked Taylor Swift's \emph{1989}?'. To answer such questions we must either wade through huge volumes of consumer reviews hoping to find one that is relevant, or otherwise pose our question directly to the community via a Q/A system.

In this paper we hope to fuse these two paradigms: given a large volume of previously answered queries about products, we hope to automatically learn whether a review of a product is relevant to a given query. We formulate this as a machine learning problem using a mixture-of-experts-type framework---here each review is an `expert' that gets to vote on the response to a particular query; simultaneously we learn a relevance function such that `relevant' reviews are those that vote correctly. At test time this learned relevance function allows us to surface reviews that are relevant to new queries on-demand. We evaluate our system, \modelname{}, on a novel corpus of 1.4 million questions (and answers) and 13 million reviews. We show quantitatively that it is effective at addressing both binary and open-ended queries, and qualitatively that it surfaces reviews that human evaluators consider to be relevant.
\end{abstract}

\section{Introduction}

Consumer reviews are invaluable as a source of data to help people form opinions on a wide range of products. Beyond telling us whether a product is `good' or `bad', reviews tell us about a wide range of \emph{personal experiences}; these include objective descriptions of the products' properties, subjective qualitative assessments, as well as unique use- (or failure-) cases.

The value and diversity of these opinions raises two questions of interest to us: (1) How can we help users navigate massive volumes of consumer opinions in order to find those that are \emph{relevant} to their decision? And (2) how can we address specific \emph{queries} that a user wishes to answer in order to evaluate a product?

To help users answer specific queries, review websites like \emph{Amazon} offer community-Q/A systems that allow users to pose product-specific questions to other consumers.\footnote{E.g.~\texttt{amazon.com/ask/questions/asin/B00B71FJU2}} Our goal here is to respond to such queries automatically and on-demand. To achieve this we make the basic insight that our two goals above naturally complement each other: given a large volume of community-Q/A data (i.e., questions and answers), and a large volume of reviews, we can automatically \emph{learn} what makes a review relevant to a query.

We see several reasons why reviews might be a useful source of information to address product-related queries, especially compared to existing work that aims to solve Q/A-like tasks by building knowledge bases of facts about the entities in question:
\begin{itemize}
 \item General question-answering is a challenging open problem. It is certainly hard to imagine that a query such as ``Will this baby seat fit in the overhead compartment of a 747?'' could be answered by building a knowledge-base using current techniques. However it is more plausible that some review of that product will contain information that is relevant to this query. By casting the problem as one of surfacing relevant opinions (rather than necessarily generating a conclusive answer), we can circumvent this difficulty, allowing us to handle complex and arbitrary queries.
 \item Fundamentally, many of the questions users ask on review websites will be those that \emph{can't} be answered using knowledge bases derived from product specifications, but rather their questions will be concerned with subjective personal experiences. Reviews are a natural and rich source of data to address such queries.
 \item Finally, the massive volume and range of opinions makes review systems difficult to navigate, especially if a user is interested in some niche aspect of a product. Thus a system that identifies opinions relevant to a specific query is of fundamental value in helping users to navigate such large corpora of reviews.
\end{itemize}

To make our objectives more concrete, we aim to formalize the problem in terms of the following goal:
\begin{quote}
 \emph{Goal:} Given a query about a particular product, we want to determine how relevant each review of that product is to the query, where `relevance' is measured in terms of how helpful the review will be in terms of identifying the correct response.
\end{quote}

\begin{figure}[t]
\begin{framed}
 \begin{center}
 \textbf{Product:}~BRAVEN BRV-1 Wireless Bluetooth Speaker
 \adjustbox{trim={.0\width} {.15\height} {0.\width} {.2\height},clip}%
  {\includegraphics[width=0.5\linewidth]{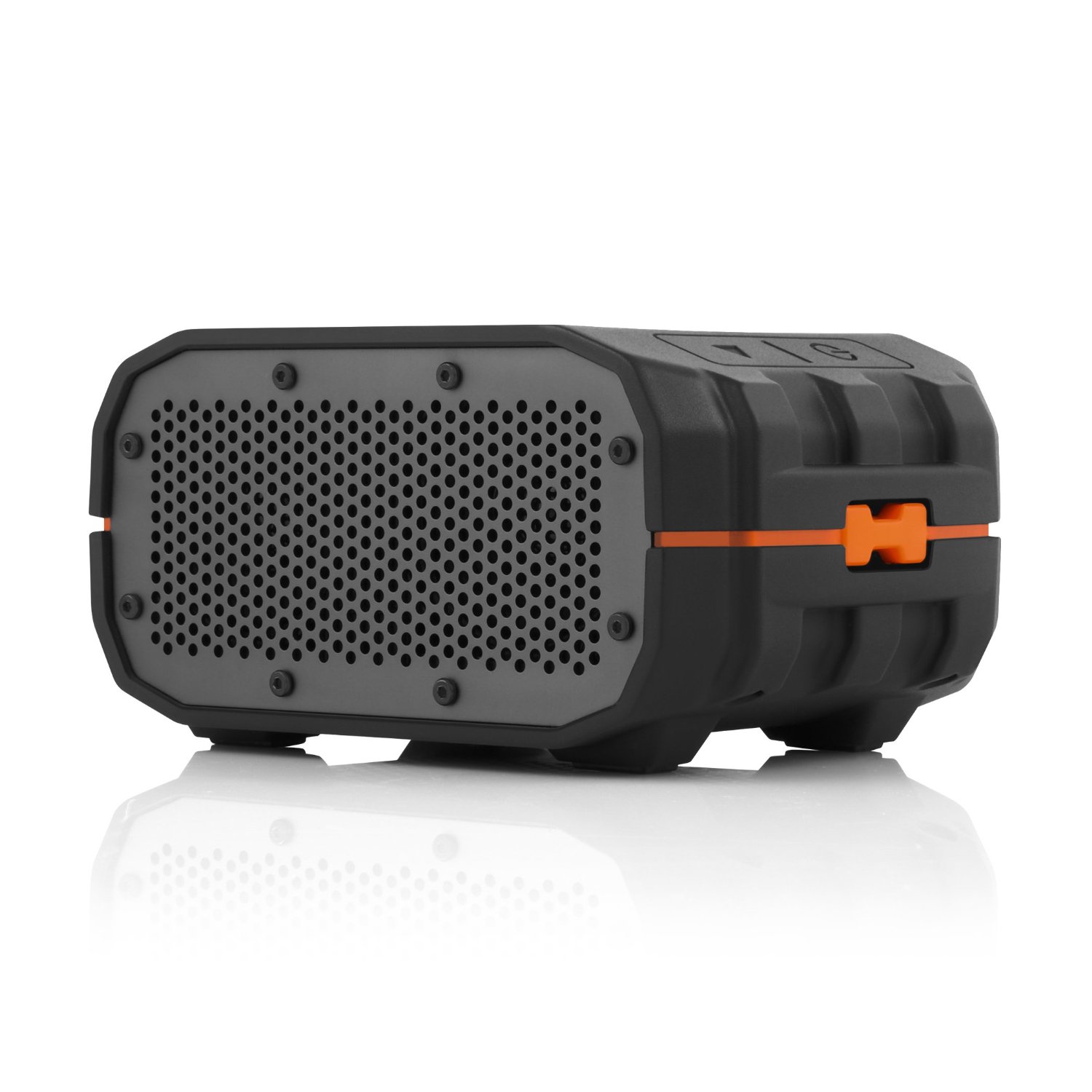}}\\
 \textbf{Query:}~``I want to use this with my iPad air while taking a jacuzzi bath. Will the volume be loud enough over the bath jets?''\\
 
 \ \\
   \begin{tabular}{m{0.78\linewidth}>{\centering\arraybackslash}m{0.07\linewidth}}
   \parbox{\linewidth}{\centering customer opinions, ranked by relevance:} & vote:\\
   \midrule
 ``The sound quality is great, especially for the size, and if you place the speaker on a hard surface it acts as a sound board, and the bass really kicks up.'' & yes\\
 \\[-2mm]
   ``If you are looking for a water resistant blue tooth speaker you will be very pleased with this product.'' & yes\\
 \\[-2mm]
   ``However if you are looking for something to throw a small party this just doesnt have the sound output.'' & no\\
 \\[-2mm]
   etc. & etc.\\
   \end{tabular}
   
   \ \\[3mm]
   \textbf{Response:} Yes
   \vspace{-3mm}
 \end{center}
\end{framed}
 \caption{An example of how our system, \modelname{}, is used. This is a real output produced by \modelname{}, given the customer query about the product above. We simultaneously learn which customer opinions are `relevant' to the query, as well as a prediction function that allows each opinion to `vote' on the response, in proportion to its relevance. These relevance and prediction functions are learned automatically from large corpora of training queries and reviews.\label{fig:examp}}
\end{figure}

The type of system we produce to address this goal is demonstrated in Figure \ref{fig:examp}. Here we surface opinions that are identified as being `relevant' to the query, which can collectively vote (along with all other opinions, in proportion to their relevance) to determine the response to the query.

This simple example demonstrates exactly the features that make our problem interesting and difficult: First, the query (`is this loud enough?') is inherently subjective, and depends on personal experience; it is hard to imagine that any fact-based knowledge repository could provide a satisfactory answer. Secondly, it is certainly a `long-tail' query---it would be hard to find relevant opinions among the (300+) reviews for this product, so a system to automatically retrieve them is valuable. Third, it is linguistically complex---few of the important words in the query appear among the most relevant reviews (e.g.~`jacuzzi bath'/`loud enough')---this means that existing solutions based on word-level similarity are unlikely to be effective. This reveals the need to learn a complex definition of `relevance' that is capable of accounting for subtle linguistic differences such as synonyms.

Finally, in the case of Figure \ref{fig:examp}, our model is able to respond to the query (in this instance correctly) with a binary answer. More importantly though, the opinions surfaced allow the user to determine the answer themselves---in this way we can extend our model to handle general open-ended queries, where the goal is not to answer the question \emph{per se}, but rather to surface relevant opinions that will help the questioner form their own conclusion.

It seems then that to address our goal we'll need a system with two components: (1) A \emph{relevance} function, to determine which reviews contain information relevant to a query, and (2) a prediction function, allowing relevant reviews to `vote' on the correct answer.

However as we stated, our main goal is \emph{not} to answer questions directly but rather to surface relevant opinions that will help the user answer the question themselves; thus it may seem as though this `voting' function is not required. Indeed, at \emph{test} time, only the relevance function is required---this is exactly the feature that shall allow our model to handle arbitrary, open-ended, and subjective queries. However the voting function is critical at \emph{training} time, so that with a large corpus of already-answered questions, we can simultaneously learn relevance and voting functions such that `relevant' reviews are those that vote for the correct answer.

The properties that we want above are captured by a classical machine learning framework known as \emph{mixtures of experts} \cite{moe}. Mixtures of experts are traditionally used when one wishes to combine a series of `weak learners'---there the goal is to simultaneously estimate (a) how `expert' each predictor is with respect to a particular input and (b) the parameters of the predictors themselves. This is an elegant framework as it allows learners to `focus' on inputs that they are good at classifying---it doesn't matter if they sometimes make incorrect predictions, so long as they correctly classify those instances where they are predicted to be experts.

In our setting, individual reviews or opinions are treated as experts that get to vote on the answer to each query; naturally some opinions will be unrelated to some queries, so we must also learn how relevant (i.e., expert) each opinion is with respect to each query. Our prediction (i.e., voting) function and relevance function are then learned simultaneously such that `relevant' opinions are precisely those that are likely to vote correctly. At test time, the relevance function can be used directly to surface relevant opinions.

We evaluate our model using a novel corpus of questions and answers from \emph{Amazon}. We consider both binary questions (such as the example in Figure \ref{fig:examp}), and open-ended questions, where reviews must vote amongst alternative answers. Quantitatively, we compare our technique to state-of-the-art methods for relevance ranking, and find that our learned definition of relevance is more capable of resolving queries compared to hand-crafted relevance measures.

Qualitatively, we evaluate our system by measuring whether human evaluators agree with the notion of `relevance' that we learn. This is especially important for open-ended queries, where it is infeasible to answer questions directly, but rather we want to surface opinions that are helpful to the user.

\subsection{Contributions}

We summarize our contributions as follows: First, we develop a new method, \modelname{}, that is able to uncover opinions that are relevant to product-related queries, and to learn this notion of relevance from training data of previously answered questions. Second, we collect a large corpus of 1.4 million answered questions and 13 million reviews on which to train the model. Ours is among the first works to combine community Q/A and review data in this way, and certainly the first to do it at the scale considered here. Third, we evaluate our system against state-of-the-art approaches for relevance ranking, where we demonstrate (a) the need to learn the notion of `relevance' from training data; (b) the need to handle heterogeneity between questions, reviews, and answers; and (c) the value of opinion data to answer product-related queries, as opposed to other data like product specifications.

Code and data is available on the first author's webpage.

\section{Related Work}

The most closely related branches of work to ours are (1) those that aim to mine and summarize opinions and facets from documents (especially from review corpora), and (2) those that study Q/A systems in general. To our knowledge our work is among the first at the interface between these two tasks, i.e., to use consumer reviews as a means of answering general queries about products, though we build upon ideas from several related areas.

\xhdr{Document summarization.} Perhaps most related to our goal of selecting relevant opinions among large corpora of reviews is the problem of \emph{multi-document summarization} \cite{mds,mds2}. Like ours, this task consists of finding relevant or `salient' parts of documents \cite{mds,chali} and intelligently combining them. Most related are approaches that apply document summarization techniques to `evaluative text' (i.e., reviews), in order to build an overview of opinions or product features \cite{mds_reviews,review_summ2,review_query3}. In contrast to our contribution, most of the above work is not `query-focused,' e.g.~the goal is to summarize product features or positive vs.~negative opinions, rather than to address specific queries, though we note a few exceptions below.

\xhdr{Relevance ranking.} A key component of the above line of work is to learn whether a document (or a phrase within a document) is relevant to a given query. `Relevance' can mean many things, from the `quality' of the text \cite{agichtein_quality}, to its lexical salience \cite{lexrank}, or its diversity compared to already-selected documents \cite{mds_reviews}. In query-focused settings, one needs a query-specific notion of relevance, i.e.,~to determine whether a document is relevant in the context of a given query. For this task, simple (yet effective) word-level similarity measures have been developed, such as Okapi BM25, a state-of-the-art TF-IDF-based relevance ranking measure \cite{bm25,bm25_plus}. A natural limitation one must overcome though is that queries and documents may be linguistically heterogeneous, so that word-level measures may fail \cite{chasm,qa_cikm14}. This can be addressed by making use of grammatical rules and phrase-level approaches (e.g.~ROUGE measures \cite{rouge}), or through probabilistic language models ranging from classical methods \cite{ponte} to recent approaches based on deep networks \cite{severyn,socher_qa}.
We discuss ranking measures more in Section \ref{sec:measures}.

\xhdr{Opinion mining.}
Studying consumer opinions, especially through rating and review datasets is a broad and varied topic. Review text has been used to augment `traditional' recommender systems by finding the aspects or facets that are relevant to people's opinions \cite{recsysJulian,wang2010,ganu} and, more related to our goal, to find `helpful' reviews \cite{bian2009learning,cdnm2} or experts on particular topics \cite{pal11}. There has also been work on generating summaries of product features \cite{huliu}, including work using multi-document summarization as mentioned above \cite{mds_reviews,review_summ2,review_query3}. This work is related in terms of the data used, and the need to learn some notion of `relevance,' though the goal is not typically to address general queries as we do here. We are aware of relatively little work that attempts to combine question-answering with opinion mining, though a few exceptions include \cite{nazi}, which answers certain types of queries on \emph{Amazon} data (e.g.~``find 100 books with over 200 5-star ratings''); or \citep{yuSubjective} which learns to distinguish `facts' from subjective opinions; or \cite{cheng_sigir15}, which tries to solve cold-start problems by finding opinion sentences of old products that will be relevant to new ones. Though in none of these cases is the goal to address general queries.

\xhdr{Q/A systems.} Many of the above ideas from multi-document summarization, relevance ranking, and topical expert-finding have been adapted to build state-of-the-art automated Q/A systems. First is `query-focused' summarization \cite{chali,query_focused}, which is similar to our task in that phrases must be selected among documents that match some query, though typically the relevance function is not learned from training data as it is here. Next (as mentioned above) is the notion that questions, answers, and documents are heterogeneous, meaning that simple bag-of-words type approaches may be insufficient to compare them \cite{chasm,qa_cikm14}, so that instead one must decompose questions \cite{random_walk} or model their syntax \cite{Moschitti07exploitingsyntactic}. Also relevant is the problem of identifying experts \cite{experts_qa,experts_qa2,jurczyk2007discovering,pal11b} or high-quality answers \cite{anderson}, or otherwise identifying instances where similar questions have already been answered elsewhere \cite{similarQ,murily}, though these differ from our paradigm in that the goal is to select among answers (or answerers), rather than to address the questions themselves.

Naturally also relevant is the large volume of Q/A work from the information retrieval community (e.g.~TREC Q/A\footnote{\url{http://trec.nist.gov/tracks.html}}); however note first that due to the data involved (in particular, subjective opinions) our approach is quite different from systems that build knowledge bases (e.g.~systems like Watson \cite{watson}), or generally systems whose task is to retrieve a list of objective facts that conclusively answer a query. Rather, our goal is to use Q/A data as a means of learning a `useful' relevance function, and as such our experiments mainly focus on state-of-the-art relevance ranking techniques.

\subsection{Key differences}

Though related to the above areas, our work is novel in a variety of ways. Our work is among the first at the interface of Q/A and opinion mining, and is novel in terms of the combination of data used, and in terms of scale. In contrast to the above work on summarization and relevance ranking, given a large volume of answered queries and a corpus of weakly relevant documents (i.e., reviews of the product being queried), our goal is to be as agnostic as possible to the definition of ``what makes an opinion relevant to a query?,'' and to learn this notion automatically from data. This also differentiates our work from traditional Q/A systems as our goal is not to answer queries directly (i.e., to output `facts' or factoids), but rather to learn a relevance function that will help users effectively navigate multiple subjective viewpoints and personal experiences. Critically, the availability of a large training corpus allows us to learn complex mappings between questions, reviews, and answers, while accounting for the heterogeneity between them.

\begin{table}[t]
 \caption{Notation. \label{tab:notation}}
\begin{center}
\begin{tabular}{lp{0.67\linewidth}}
\toprule
Symbol & Description\\
\midrule
$q \in \mathcal Q$, $a \in \mathcal A$ & query and query set, answer and answer set\\
$y \in \mathcal Y$ & label set (for binary questions)\\
$r \in \mathcal R$ & review and review set\\
$s$ & relevance/scoring function\\
$v$ & prediction/voting function\\
$\delta$ & indicator function ($1$ iff the argument is true)\\
$\theta$, $\vartheta,A,B$ & terms in the bilinear relevance function\\
$\vartheta',X,Y$ & terms in the bilinear prediction function\\
$p(r|q)$ & relevance of a review $r$ to a query $q$\\
$p(y|r,q)$ & probability of selecting a positive answer to a query $q$ given a review $r$\\
$p(a > \bar{a} | r)$ & preference of answer $a$ over $\bar{a}$\\
\bottomrule
\end{tabular}
\end{center}
\end{table}

\section{Model preliminaries}
\label{sec:prelim}

Since our fundamental goal is to learn relevance functions so as to surface useful opinions in response to queries, we mainly build upon and compare to existing techniques for relevance ranking. We also briefly describe the mixture-of-experts framework (upon which we build our model) before we describe \modelname{} in Section \ref{sec:square}.

\subsection{Standard measures for relevance ranking}
\label{sec:measures}

We first describe a few standard measures for relevance ranking, given a query $q$ and a document $d$ (in our case, a question and a review), whose relevance to the query we want to determine.

\xhdr{Cosine similarity} is a simple similarity measure that operates on Bag-of-Words representations of a document and a query. Here the similarity is given by
\begin{equation}
 \mathit{cos}(q,d) = \frac{q \cdot d}{\| q\| \| d \|},
 \label{eq:cosine}
\end{equation}
i.e., the cosine of the angle between (the bag-of-words representations of) the query $q$ and a document $d$. This can be further refined by weighting the individual dimensions, i.e.,
\begin{equation}
 \mathit{cos}_\vartheta(q,d) = \frac{(q \odot d) \cdot \theta}{\| q\| \| d \|},
 \label{eq:cosineL}
\end{equation}
where $(q \odot d)$ is the Hadamard product.

\xhdr{Okapi BM25} is state-of-the-art among `TF-IDF-like' ranking functions and is regularly used for document retrieval tasks \citep{bm25,irbook}. TF-IDF-based ranking measures address a fundamental issue with measures like the cosine similarity (above) whereby common---but irrelevant---words can dominate the ranking function. This can be addressed by defining a ranking function that rewards words which appear many times in a selected document (high TF), but which are rare among other documents (high IDF). Okapi BM25 is a parameterized family of functions based on this idea:
\begin{equation}
 \mathit{bm25}(q, d) = \sum_{i=1}^n \frac{\text{IDF}(q_i) \cdot f(q_i,d)\cdot(k_1 + 1)}{f(q_i,d) + k_1 \cdot (1 - b + b \cdot \frac{|d|}{\text{avgdl}})}.
 \label{eq:bm25}
\end{equation}
Again $q$ and $d$ are the query and a document, and $f$ and $\text{IDF}$ are the term frequency (of a word $q_i$ in the query) and inverse document frequency as described above. `$\text{avgdl}$' is the average document length, and $b$ and $k_1$ are tunable parameters, which we set as described in \cite{irbook}. See \cite{bm25,irbook} for further detail.

Essentially, we treat BM25 as a state-of-the-art `off-the-shelf' document ranking measure that we can use for evaluation and benchmarking, and also as a feature for ranking in our own model.

\xhdr{Bilinear models.} While TF-IDF-like measures help to discover rare but important words, an issue that still remains is that of \emph{synonyms}, i.e., different words being used to refer to the same concept, and therefore being ignored by the similarity measure in question. This is especially an issue in our setting, where questions and reviews are only tangentially related and may draw from very different vocabularies \cite{chasm,qa_cikm14}---thus one needs to learn that a word used in (say) a question about whether a baby seat fits in overhead luggage is `related to' a review that describes its dimensions.

Bilinear models \cite{chuBilinear,bilinear0,bilinear} can help to address this issue by learning complex mappings between words in one corpus and words in another (or more generally between arbitrary feature spaces). Here compatibility between a query and a document is given by
\begin{equation}
 qMd^T = \sum_{i,j} M_{ij} q_i d_j,
\end{equation}
where $M$ is a matrix whose entry $M_{ij}$ encodes the relationship between a term $q_i$ in the query and a term $d_j$ in the document (setting $M=I$ on normalized vectors recovers the cosine similarity). This is a highly flexible model, which even allows that the dimensions of the two feature spaces be different; in practice, since $M$ is very high-dimensional (in our application, the size of the vocabulary squared), we assume that it is low-rank, i.e., that it can be approximated by $M \sim AB^T$ where $A$ and $B$ are each rank $K$.\footnote{This is similar to the idea proposed by Factorization Machines \cite{rendle_fm}, allowing complex pairwise interactions to be handled by assuming that they have low-rank structure (i.e., they factorize).} Thus our similarity measure becomes
\begin{equation}
 qAB^Td^T = (qA) \cdot (dB).
\end{equation}
This has an intuitive explanation, which is that $A$ and $B$ project terms from the query and the document into a low-dimensional space such that `similar' terms (such as synonyms) in the query and the document are projected nearby (and have a high inner product).

\subsection{Mixtures of Experts}
\label{sec:moe}

\emph{Mixtures of experts} (MoEs) are a classical way to combine the outputs of several classifiers (or `weak learners') by associating weighted confidence scores with each classifier \cite{moe}. In our setting `experts' shall be individual reviews, each of which lends support for or against a particular response to a query. The value of such a model is that relevance and classification parameters are learned \emph{simultaneously}, which allows individual learners to focus on classifying only those instances where they are considered `relevant,' without penalizing them for misclassification elsewhere. In the next section we show how this is useful in our setting, where only a tiny subset of reviews may be helpful in addressing a particular query.

Generally speaking, for a binary classification task, each expert outputs a probability associated with a positive label. The final classification output is then given by aggregating the predictions of the experts, in proportion to their confidence (or expertise). This can be expressed probabilistically as
\begin{equation}
 p(y | X) = \sum_f \overbrace{p(f | X)}^{\mathclap{\text{confidence in $f$'s ability to classify $X$}}} \underbrace{p(y | f, X)}_{\mathclap{\text{$f$'s prediction}}}.
 \label{eq:moe}
\end{equation}

Here our confidence in each expert, $p(f|X)$, is treated as a probability, which can be obtained from an arbitrary real-valued score $s(f,X)$ using a softmax function:
\begin{equation}
 p(f | X) = \frac{\exp(s(f,X))}{\sum_{f'}\exp(s(f',X))}.
\end{equation}
Similarly for binary classification tasks the prediction of a particular expert can be obtained using a logistic function:
\begin{equation}
 p(y | f, X) = \sigma(v(f,X)) = \frac{1}{1 + e^{-v(f,X)}}.
 \label{eq:logistic}
\end{equation}
Here $s$ and $v$ are our `relevance' and `voting' functions respectively. To define an MoE model, we must now define (parameterized) functions $s(f,X)$ and $v(f,X)$, and tune their parameters to maximize the likelihood of the available training labels. We next describe how this formulation can be applied to queries and reviews, and describe our parameter learning strategy in Section \ref{sec:learning}.

\section{MOQA}
\label{sec:square}

We now present our model, \emph{\modelnamelong{}}, or \modelname{} for short. In the previous section we outlined the `Mixture of Experts' framework, which combines weak learners by aggregating their outputs with weighted confidence scores. Here, we show that such a model can be adapted to simultaneously identify relevant reviews, and combine them to answer complex queries, by treating reviews as experts that either support or oppose a particular response.

\subsection{Mixtures of Experts for review relevance ranking}

As described in Section \ref{sec:moe}, our MoE model is defined in terms of two parameterized functions: $s$, which determines whether a review (`expert') is relevant to the query, and $v$, which given the query \emph{and} a review makes a prediction (or vote). Our goal is that predictions are correct exactly for those reviews considered to be relevant. We first define our relevance function $s$ before defining our prediction functions for binary queries in Section \ref{sec:qbinary} and arbitrary queries in Section \ref{sec:qarbitrary}.

Our scoring function $s(r, q)$ defines the relevance of a review $r$ to a query $q$. In principle we could make use of any of the relevance measures from Section \ref{sec:measures} `as is,' but we want our scoring function to be \emph{parameterized} so that we can learn from training data what constitutes a `relevant' review. Thus we define a parameterized scoring function as follows:
\begin{equation}
 s_{\Theta}(r,q) = \underbrace{\phi(r,q) \cdot \theta}_{{\text{pairwise similarity}\vphantom{ly}}} + \underbrace{\psi(q) M \psi(r)^T}_{\text{bilinear model}\vphantom{ly}}.
 \label{eq:relevance}
\end{equation}
Here $\phi(r,q)$ is a feature vector that is made up of existing pairwise similarity measures. $\theta$ then weights these measures so as to determine how they should be combined in order to achieve the best ranking. Thus $\phi(r,q)$ allows us to straightforwardly make use of existing `off-the-shelf' similarity measures that are considered to be state-of-the-art. In our case we make use of BM25+ \cite{bm25_plus} and ROUGE-L \cite{rouge} (longest common subsequence) features, though we describe our experimental setup in more detail in Section \ref{sec:experiments}.

The second expression in \eq{eq:relevance} is a bilinear scoring function between features of the query ($\psi(q)$) and the review ($\psi(r)$). As features we us a simple bag-of-words representation of the two expressions with an $F = 5000$ word vocabulary. As we suggested previously, learning an $F \times F$ dimensional parameter $M$ is not tractable, so we approximate it by
\begin{equation}
M = \underbrace{(\psi(q) \odot \psi(r)) \cdot \vartheta}_{\text{diagonal term}} + \underbrace{\psi(q) AB^T \psi(r)^T}_{\text{low-rank term}}.
\label{eq:approxtransform}
\end{equation}
$\vartheta$ (the diagonal component of $M$) then accounts for simple term-to-term similarity, whereas $A$ and $B$ (the low-rank component of $M$) are projections that map $\psi(q)$ and $\psi(r)$ (respectively) into $K$-dimensional space ($K=5$ in our experiments) in order to account for linguistic differences (such as synonym use) between the two sources of text. Thus rather than fitting $F \times F$ parameters  we need to fit only $(2K + 1)\cdot F$ parameters in order to approximate $M$.

To obtain the final relevance function, we optimize all parameters $\Theta = \lbrace \theta,\vartheta,A,B \rbrace$ using supervised learning, as described in the following section.

\subsection{Binary (i.e., yes/no) questions}
\label{sec:qbinary}
\label{sec:learning}

Dealing with binary (yes/no) questions is a relatively straightforward application of an MoE-type model, where each of the `experts' (i.e., reviews) must make a binary prediction as to whether the query is supported by the content of the review. This we also achieve using a bilinear scoring function:
\begin{equation}
 v_{\Theta'}(q,r) = (\psi(q) \odot \psi(r)) \cdot \vartheta' + \psi(q) XY^T \psi(r)^T.
 \label{eq:prediction}
\end{equation}
Note that this is different from the relevance function $s$ in \eq{eq:relevance} (though it has a similar form). The role of \eq{eq:prediction} above is to vote on a binary outcome; how much weight/relevance is given to this vote is determined by \eq{eq:relevance}. Positive/negative $v(q,r)$ corresponds to a vote in favor of a positive or negative answer (respectively).

\xhdr{Learning.}
Given a training set of questions with labeled yes/no answers (to be described in Section \ref{sec:expbinary}), our goal is to optimize the relevance parameters $\Theta = \lbrace \theta,\vartheta,A,B \rbrace$ and the prediction parameters $\Theta' = \lbrace \vartheta', X, Y \rbrace$ simultaneously so as to maximize the likelihood that the training answers will be given the correct labels. In other words, we want to define these functions such that reviews given high relevance scores are precisely those that help to predict the correct answer. Using the expression in \eq{eq:moe}, the likelihood function is given by
\begin{equation}
 L_{\Theta,\Theta'}(\mathcal Y|\mathcal Q, \mathcal R) = \!\!\prod_{q \in \mathcal Q^{(\text{train})}_{\text{yes}}}\!\! p_{\Theta,\Theta'}(y | q) \!\!\prod_{q \in \mathcal Q^{(\text{train})}_{\text{no}}}\!\! (1 - p_{\Theta,\Theta'}(y|q)),
 \label{eq:likelihood}
\end{equation}
where $\mathcal Q^{(\text{train})}_{\text{yes}}$ and $\mathcal Q^{(\text{train})}_{\text{no}}$ are training sets of questions with positive and negative answers, and $\mathcal Y$ and $\mathcal R$ are the label set and reviews respectively. $p(y|q)$ (the probability of selecting the answer `yes' given the query $q$) is given by
\begin{equation}
 p_{\Theta,\Theta'}(y|q) = \sum_{r \in \mathcal R_{i(q)}} \biggl\lbrace \underbrace{\frac{e^{s_\Theta(q,r)}}{\sum_{r' \in \mathcal R_{i(q)}} e^{s_\Theta(q,r')}}}_{\text{relevance}} \underbrace{\frac{1}{1 + e^{-v_{\Theta'}(q,r)}}}_{\text{prediction}} \biggr\rbrace,
 \label{eq:binarymodel}
\end{equation}
where $\mathcal R_{i(q)}$ is the set of reviews associated with the item referred to in the query $q$. We optimize the (log) likelihood of the parameters in \eq{eq:likelihood} using L-BFGS, a quasi-Newton method for non-linear optimization of problems with many variables. We added a simple $\ell_2$ regularizer to the model parameters, though did not run into issues of overfitting, as the number of parameters is far smaller than the number of samples available for training.

\subsection{Open-ended questions}
\label{sec:qarbitrary}

While binary queries already account for a substantial fraction of our dataset, and are a valuable testbed for quantitatively evaluating our method, we wish to extend our method to arbitrary open-ended questions, both to increase its coverage, and to do away with the need for labeled yes/no answers at training time.

Here our goal is to train a method that given a corpus of candidate answers (one of which is the `true' answer that a responder provided) will assign a higher score to the true answer than to all non-answers. Naturally in a live system one does not have access to such a corpus containing the correct answer, but recall that this is not required: rather, we use answers only at \emph{training} time to learn our relevance function, so that at test time we can surface relevant reviews \emph{without} needing candidate answers to be available.

Specifically, we want to train the model such that the true answer is given a higher rank than all non-answers, i.e., to train a ranking function to maximize the average Area Under the Curve (AUC):
\begin{equation}
 \mathit{AUC}^{(\text{train})} = \frac{1}{|\mathcal Q^{(\text{train})}|}\sum_{q\in \mathcal Q^{(\text{train})}} \frac{1}{|\mathcal A|} \sum_{\bar{a} \in \mathcal A} \delta(a(q) > \bar{a}),
\end{equation}
where $a(q)$ is the `true' answer for the query $q$ ($\mathcal A$ is the answer set) and $\delta(a(q) > \bar{a})$ is an indicator counting whether this answer was preferred over a non-answer $\bar{a}$. In other words, the above simply counts the fraction of cases where the true answer was considered better than non-answers.

In practice, the AUC is (approximately) maximized by optimizing a pairwise ranking measure, where the true answer should be given a higher score than a (randomly chosen) non-answer, i.e., instead of optimizing $p_{\Theta,\Theta'}(y|q)$ from \eq{eq:binarymodel} we optimize
$$
 p(a > \bar{a}|q) \sum_r \overbrace{p(r | q)}^{\mathclap{\text{relevance}}} \underbrace{p(a > \bar{a}|r)}_{\mathclap{\text{$a$ is a better answer than $\bar{a}$}}}.
$$
To do so we make use of the same relevance function $s$ and the same scoring function $v$ used in \eq{eq:prediction}, with two important differences: First, the scoring function takes a candidate answer (rather than the query) as a parameter (i.e., $v(a,r)$ rather than $v(q,r)$). This is because our goal is no longer to estimate a binary response to the query $q$, but rather to determine whether the answer $a$ is supported by the review $r$. Second, since we want to use this function to rank answers, we no longer care that $v(a,r)$ is maximized, but rather that $v(a,r)$ (for the \emph{true} answer) is higher than $v(\bar{a},r)$ for non-answers $\bar{a}$. This can be approximated by optimizing the logistic loss
\begin{equation}
 p(a > \bar{a}|r) = \sigma(v(a,r) - v(\bar{a},r)) = \frac{1}{1 + e^{v(\bar{a},r) - v(a,r)}}.
\end{equation}
This will approximate the AUC if enough random non-answers are selected; optimizing pairwise ranking losses as a means of optimizing the AUC is standard practice in recommender systems that make use of implicit feedback \cite{bpr}.
Otherwise, training proceeds as before, with the two differences being that (1) $p(a > \bar{a}|r)$ replaces the prediction function in \eq{eq:binarymodel}, and (2) multiple non-answers must be sampled for training. In practice we use 10 epochs (i.e., we generate 10 random non-answers per query during each training iteration). On our largest dataset (\emph{electronics}), training requires around 4-6 hours on a standard desktop machine.

\section{Experiments}
\label{sec:experiments}

We evaluate \modelname{} in terms of three aspects: First for binary queries, we evaluate its ability to resolve them. Second, for open-ended queries, its ability to select the correct answer among alternatives. Finally we evaluate \modelname{} qualitatively, in terms of its ability to identify reviews that humans consider to be relevant to their query. We evaluate this on a large dataset of reviews and queries from \emph{Amazon}, as described below.

\subsection{Data}

We collected review and Q/A data from \emph{Amazon.com}. We started with a previous crawl from \cite{McAPanLes15}, which contains a snapshot of product reviews up to July 2014 (but which includes only review data). For each product in that dataset, we then collected all questions on its Q/A page, and the top-voted answer chosen by users. We also crawled descriptions of all products, in order to evaluate how description text compares to text from reviews. This results in a dataset of 1.4 million questions (and answers) on 191 thousand products, about which we have over 13 million customer reviews. We train separate models for each top-level category (electronics, automotive, etc.). Statistics for the 8 largest categories (on which we report results) are shown in Table \ref{tab:data}.

\begin{table}
\caption{Dataset Statistics.\label{tab:data}}
\renewcommand{\tabcolsep}{2pt}
\begin{center}
\begin{tabular}{lrrr}
\toprule
Dataset & \parbox{0.21\linewidth}{\centering questions\\ (w/ answers)}\hspace{-2.5mm} & \ \ \ \ products & reviews\\
\midrule
electronics                                        & 314,263 & 39,371 & 4,314,858 \\
home and kitchen                                   & 184,439 & 24,501 & 2,012,777 \\
sports and outdoors                                & 146,891 & 19,332 & 1,013,196 \\
tools and home impr.                               & 101,088 & 13,397 & 752,947 \\
automotive                                         & 89,923 & 12,536 & 395,872 \\
cell phones                                        & 85,865 & 10,407 & 1,534,094 \\
health and personal care                           & 80,496 & 10,860 & 1,162,587 \\
patio lawn and garden                              & 59,595 & 7,986 & 451,473 \\
\midrule
total                                              & 1,447,173 & 191,185 & 13,498,681 \\
\bottomrule
\end{tabular}
\end{center}
\renewcommand{\tabcolsep}{6pt}
\end{table}

\subsection{Labeling yes/no answers}
\label{sec:expbinary}

Although the above data is already sufficient for addressing open-ended questions, for binary questions we must first obtain additional labels for training. Here we need to identify whether each question in our dataset is a yes/no question, and if so, whether it has a yes/no answer. In spite of this need for additional labels, addressing yes/no questions is valuable as it gives us a simple and objective way to evaluate our system.

We began by manually labeling one thousand questions to identify those which were binary, and those which had binary answers (note that these are not equivalent concepts, as some yes/no questions may be answered ambiguously). We found that 56.1\% of questions are binary, and that 76.5\% of these had conclusive binary answers. Of those questions with yes/no answers, slightly over half (62.4\%) had positive (i.e., `yes') answers.

Note that the purpose of this small, manually labeled sample is not to train \modelname{} but rather to evaluate simple techniques for automatically labeling yes/no questions and answers. This is much easier than our overall task, since we are \emph{given} the answer and simply want to determine whether it was positive or negative, for which simple NLP techniques suffice. 

To identify whether a question is binary, a recent approach developed by \emph{Google} proved to be effective \cite{yesno}. This approach consists of a series of complex grammatical rules which are used to form regular expressions, which essentially identify occurrences of `be', modal, and auxiliary verbs. Among our labeled data these rules identified yes/no questions with 97\% precision at 82\% recall.
Note that in this setting we are perfectly happy to sacrifice some recall for the sake of precision---what we want is a sufficiently large sample of labeled yes/no questions to train \modelname{}, but we are willing to discard ambiguous cases in order to do so.

Next we want to label \emph{answers} as being yes/no. Ultimately we trained a simple bag-of-unigrams SVM, plus an additional feature based on the first word only (which is often simply `yes' or `no').
Again, since we are willing to sacrifice recall for precision, we discarded test instances that were close to the decision hyperplane. By keeping only the 50\% of instances about which the classifier was the most confident, we obtained 98\% classification accuracy on held-out data.

Finally we consider a question only if \emph{both} of the above tests pass, i.e., the question is identified as being binary \emph{and} the answer is classified as yes/no with high confidence. Ultimately through the above process we obtained 309,419 questions that we were able to label with high confidence, which can be used to train the binary version of \modelname{} in Section \ref{sec:expyesno}.

\subsection{Baselines}
\label{sec:baselines}

We compare \modelname{} against the following baselines:

\begin{description}
\item[rand] ranks and classifies all instances randomly. By definition this has 50\% accuracy (on average) for both of the tasks we consider. Recall also that for yes/no questions around 62\% are answered affirmatively, roughly reflecting the performance of `always yes' classification.

\item[Cosine similarity (c).] The relevance of a review to a query is determined by their cosine similarity, as in \eq{eq:cosine}.

\item[Okapi-BM25+ (o).] BM25 is a state-of-the-art TF-IDF-based relevance measure that is commonly used in retrieval applications \cite{bm25,irbook}. Here we use a recent extension of BM25 known as BM25+ \citep{bm25_plus}, which includes an additional term ($\delta \sum_{i=1}^n \text{IDF}(q_i)$) in the above expression in order to lower-bound the normalization by document length.

\item[ROUGE-L (r).] Review relevance is determined by ROUGE metrics, which are commonly used to measure similarity in document summarization tasks \cite{rouge}. Here we use ROUGE-L (longest common subsequence) scores.

\item[Learning vs.~non learning (-L).] The above measures (c), (o), and (r) can be applied `off the shelf,' i.e., without using a training set. We analyze the effect of applying maximum-likelihood training (as in eq.~\ref{eq:likelihood}) to tune their parameters (c-L, o-L, etc.).

\item[\emph{Mdqa}] is the same as \modelname{}, except that reviews are replaced by product descriptions.
\end{description}

The above baselines are designed to assess (1) the efficacy of existing state-of-the-art `off-the-shelf' relevance measures for the ranking tasks we consider (c, o, and r); (2) the benefit of using a training set to optimize the relevance and scoring functions (c-L, o-L, etc.); (3) the effectiveness of reviews as a source of data versus other potential knowledge bases (\emph{Mdqa}); and finally (4) the influence of the bilinear term and the performance of \modelname{} itself.

For the baselines above we use a linear scoring function in the predictor ($v_{\Theta'}(q,r) = (\psi(q) \odot \psi(r)) \cdot \vartheta'$), though for \emph{Mdqa} and \modelname{} we also include the bilinear term as in \eq{eq:prediction}. Recall that our model already includes the cosine similarity, ROUGE score, and BM25+ measures as features, so that comparison between the baseline `cro-L' (i.e., all of the above measures weighted by maximum likelihood) and \modelname{} essentially assesses the value of using bilinear models for relevance ranking.

For all methods, we split reviews at the level of \emph{sentences}, which we found to be more convenient when surfacing results via an interface, as we do in our qualitative evaluation. We found that this also led to slightly (but consistently) better performance than using complete reviews---while reviews contain more information, sentences are much better targeted to specific product details.

\subsection{Quantitative evaluation}

\subsubsection{Yes/no questions}
\label{sec:expyesno}

We first evaluate our method in terms of its ability to correctly classify held-out yes/no questions, using the binary groundtruth described above. Here we want to measure the classification accuracy (w.r.t.~a query set $\mathcal Q$):
\begin{multline*}
\text{accuracy}(\mathcal Q) =\\
\frac{1}{|\mathcal Q|} \sum_{q \in \mathcal Q} \underbrace{\delta(q \in \mathcal Q_{\text{yes}})\delta(p(y|q)\! >\! \frac{1}{2})}_{\text{true positives}} + \underbrace{\delta(q \in \mathcal Q_{\text{no}})\delta(p(y_q)\! <\! \frac{1}{2})}_{\text{true negatives}},
\end{multline*}
i.e., the fraction of queries that were given the correct binary label.

We found this to be an incredibly difficult measure to perform well on (for any method), largely due to the fact that some fraction of queries are simply not addressed among the reviews available. Fortunately, since we are training probabilistic classifiers, we can also associate a \emph{confidence} with each classification (i.e., its distance from the decision boundary, $|\frac{1}{2} - p(y|q)|$). Our hope is that a good model will assign high confidence scores to exactly those queries that can be (correctly) addressed. To evaluate algorithms as a function of confidence, we consider the accuracy@k:
\begin{equation}
 \text{A@k} = \text{accuracy}\biggl(\underbrace{\argmax_{\mathcal Q' \in \mathcal{P}_k(\mathcal Q)}\sum_{q \in \mathcal Q'} |\frac{1}{2} - p(y|q)|}_{k \text{\ most confidence predictions}}\biggr),
\end{equation}
Where $\mathcal{P}_k(\mathcal Q)$ is the set of $k$-sized subsets of $\mathcal Q$.

Table \ref{tab:yn} shows the performance of \modelname{} and baselines, in terms of the accuracy@50\% (i.e., for the 50\% of predictions about which each algorithms is most confident). Note that only methods with learning (-L) are shown as non-learning approaches are not applicable here (since there is no good way to determine parameters for a binary decision function in eq.~\ref{eq:binarymodel} \emph{without} learning). Here \modelname{} is substantially more accurate than alternatives, especially on larger datasets (where more data is available to learn a meaningful bilinear map). Among the baselines {ro-L} (ROUGE+Okapi BM25+ with learned weights) was the second strongest, with additional similarity-based features (cro-L) helping only slightly.

\begin{table}
 \caption{Performance of \modelname{} against baselines in terms of the accuracy@50\%; only learning (i.e., -L) baselines are shown as non-learning baselines are not applicable to this task. \label{tab:yn}}
 \begin{center}
 \renewcommand{\tabcolsep}{2.5pt}
 \begin{tabular}{lcccc|c}
 \toprule
 & \parbox{0.035\textwidth}{\centering rand\vphantom{Lq}}
 & \parbox{0.05\textwidth}{\centering {ro-L}\vphantom{Lq}}
 & \parbox{0.05\textwidth}{\centering {cro-L}\vphantom{Lq}}
 & \parbox{0.05\textwidth}{\centering \emph{Moqa}\vphantom{Lq}}
 & \parbox{0.065\textwidth}{\centering \small red.~in error\\ vs.~cro\nobreakdash-L}\\
 \midrule
 electronics                  & 50\% & 78.9\% & 79.7\% & \textbf{82.6\%} & \hphantom{0}3.7\% \\
home and kitchen              & 50\% & 70.3\% & 64.6\% & \textbf{73.6\%} & 13.9\% \\
sports and outdoors           & 50\% & 71.9\% & 72.8\% & \textbf{74.1\%} & \hphantom{0}1.8\% \\
tools and home impr.          & 50\% & 70.7\% & 69.0\% & \textbf{73.2\%} & \hphantom{0}6.1\% \\
automotive                    & 50\% & 74.8\% & 76.6\% & \textbf{78.4\%} & \hphantom{0}2.3\% \\
cell phones                   & 50\% & 74.6\% & 76.3\% & \textbf{79.4\%} & \hphantom{0}4.1\% \\
health and pers.~care      & 50\% & 61.7\% & 75.5\% & \textbf{76.2\%} & \hphantom{0}0.9\% \\
patio lawn and garden         & 50\% & 74.6\% & 75.4\% & \textbf{76.8\%} & \hphantom{0}1.8\% \\
\midrule
average & 50\% & 72.2\% & 73.7\% & \textbf{76.8\%} & \hphantom{0}4.3\% \\
\bottomrule
 \end{tabular}
 \end{center}
\renewcommand{\tabcolsep}{6pt}
\end{table}

\begin{figure}[t]
 \begin{center}
 \hspace{-.275in}
  \includegraphics{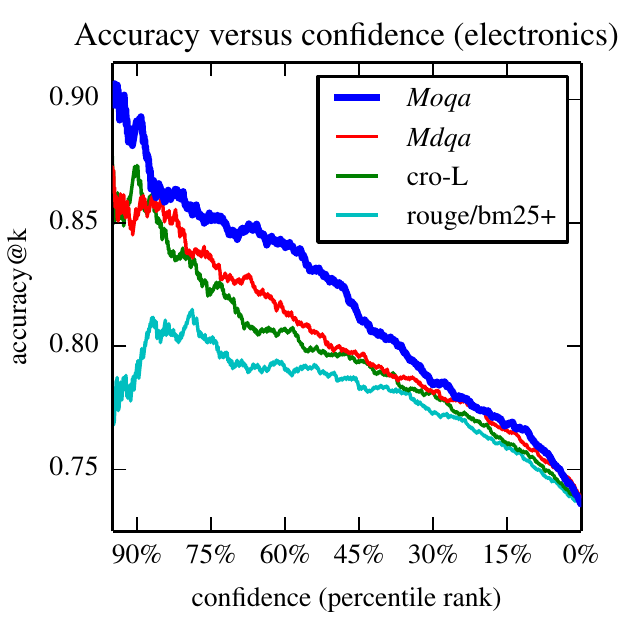}
 \end{center}
\caption{Accuracy as a function of confidence. \modelname{} correctly assigns high confidence to those queries it is able to accurately resolve.\label{fig:prec}}
\end{figure}

Figure \ref{fig:prec} shows the full spectrum of accuracy as a function of confidence on `electronics' queries, i.e., it shows how performance degrades as confidence decreases (other categories yielded similar results). Indeed we find that for all methods performance degrades for low-confidence queries. Nevertheless \modelname{} remains more accurate than alternatives across the full confidence spectrum, and for queries about which it is most confident obtains an accuracy of around 90\%, far exceeding the performance of any baseline. Figure \ref{fig:prec} also shows the performance of \emph{Mdqa}, as we discuss below.

\subsubsection{Open-ended questions}

In Table \ref{tab:openended} we show the performance of \modelname{} against baselines for open-ended queries on our largest datasets. Cosine similarity (c) was the strongest non-learning baseline, slightly outperforming the ROUGE score (r) and BM25+ (o, not shown for brevity). Learning improved all baselines, with the strongest being ROUGE and BM25+ combined (ro-L), over which adding weighted cosine similarity did not further improve performance (cro-L), much as we found with binary queries above. \modelname{} was strictly dominant on all datasets, reducing the error over the strongest baseline by 50.6\% on average.

\begin{table*}
\caption{Performance of \modelname{} against baselines (a key is shown at right for baselines from Section \ref{sec:baselines}). Reported numbers are average AUC (i.e., the models' ability to assign the highest possible rank to the correct answer); higher is better.\label{tab:openended}}

\renewcommand{\tabcolsep}{3pt}

\begin{tabular}{lccccccc|cc}
\toprule
Dataset\vphantom{Lq}
& \parbox{0.03\textwidth}{\centering rand\vphantom{Lq}}
& \parbox{0.05\textwidth}{\centering c\vphantom{Lq}}
& \parbox{0.05\textwidth}{\centering r\vphantom{Lq}}
& \parbox{0.05\textwidth}{\centering {ro-L}\vphantom{Lq}}
& \parbox{0.05\textwidth}{\centering {cro-L}\vphantom{Lq}}
& \parbox{0.05\textwidth}{\centering \emph{Mdqa}\vphantom{Lq}}
& \parbox{0.05\textwidth}{\centering \emph{Moqa}\vphantom{Lq}}
& \parbox{0.07\textwidth}{\centering \small red.~in error\\ vs.~cro\nobreakdash-L}
& \parbox{0.07\textwidth}{\centering \small red.~in error\\ vs.~\emph{Mdqa}}\\
\midrule
electronics                                      & 0.5 & 0.633 & 0.626 & 0.886 & 0.855 & 0.865 & \textbf{0.912} & 65.6\% & 54.5\% \\
home and kitchen                                 & 0.5 & 0.643 & 0.635 & 0.850 & 0.840 & 0.863 & \textbf{0.907} & 73.5\% & 48.1\% \\
sports and outdoors                              & 0.5 & 0.653 & 0.645 & 0.848 & 0.845 & 0.860 & \textbf{0.885} & 35.1\% & 22.5\% \\
tools and home impr.                             & 0.5 & 0.638 & 0.632 & 0.860 & 0.817 & 0.834 & \textbf{0.884} & 58.8\% & 43.7\% \\
automotive                                       & 0.5 & 0.648 & 0.640 & 0.810 & 0.821 & 0.825 & \textbf{0.863} & 30.4\% & 27.7\% \\
cell phones                                      & 0.5 & 0.624 & 0.617 & 0.768 & 0.797 & 0.844 & \textbf{0.886} & 78.7\% & 37.5\% \\
health and personal care                         & 0.5 & 0.632 & 0.625 & 0.818 & 0.817 & 0.842 & \textbf{0.880} & 52.7\% & 31.9\% \\
patio lawn and garden                            & 0.5 & 0.634 & 0.628 & 0.835 & 0.833 & 0.796 & \textbf{0.848} & 10.2\% & 34.4\% \\
\midrule
average                                          & 0.5 & 0.638 & 0.631 & 0.834 & 0.828 & 0.841 & \textbf{0.883} & 50.6\% & 37.5\% \\
\bottomrule
\renewcommand{\tabcolsep}{6pt}
\end{tabular}\ 
\begin{tabular}{c|l}
\vphantom{Lq} rand & random\\
\vphantom{Lq} c & cosine similarity\\
\vphantom{Lq} r & ROUGE measures\\
\vphantom{Lq} o & Okapi BM25+\\
\vphantom{Lq} -L & ML parameters\\
 \vphantom{Lq}\emph{Moqa} & our method\\
 \vphantom{Lq}\emph{Mdqa} & w/ descriptions
\end{tabular}
\end{table*}

\subsubsection{Reviews versus product descriptions}
\label{sec:descriptions}

We also want to evaluate whether review text is a better source of data than other sources, such as product descriptions or specifications. To test this we collected description/specification text for each of the products in our catalogue. From here, we simply interchange reviews with descriptions (recall that both models operate at the level of sentences). We find that while \modelname{} with descriptions (i.e., \emph{Mdqa}) performs well (on par with the strongest baselines), it is still substantially outperformed when we use review text. Here \modelname{} yields a 37.5\% reduction in error over \emph{Mdqa} in Table \ref{tab:openended}; similarly in Figure \ref{fig:prec}, for binary queries \emph{Mdqa} is on par with the strongest baseline but substantially outperformed by \modelname{} (again other datasets are similar and not shown for brevity). 

Partly, reviews perform better because we want to answer subjective queries that depend on personal experiences, for which reviews are simply a more appropriate source of data. But other than that, reviews are simply more abundant---we have on the order of 100 times as many reviews as descriptions (products with active Q/A pages tend to be reasonably popular ones); thus it is partly the sheer volume and diversity of reviews available that makes them effective as a means of answering questions.

We discuss these findings in more detail in Section \ref{sec:discussion}.

\subsection{Qualitative evaluation}

Finally, we evaluate \modelname{} qualitatively through a user study. Although we have shown \modelname{} to be effective at correctly resolving binary queries, and at maximizing the AUC to select a correct answer among alternatives, what remains to be seen is whether the relevance functions that we learned to do so are aligned with what humans consider to be `relevant.' Evaluating this aspect is especially important because in a live system our approach would presumably not be used to answer queries directly (which we have shown to be very difficult, and in general still an open problem), but rather to surface relevant reviews that will help the user to evaluate the product themselves.

Here we use the relevance functions $s_\Theta(q,r)$ that we learned in the previous section (i.e., from Table \ref{tab:openended}) to compare which definition of `relevance' is best aligned with real users' evaluations---note that the voting function $v$ is not required at this stage.

\begin{figure}[t]
 \begin{center}
  \fbox{\includegraphics[width=0.8\linewidth]{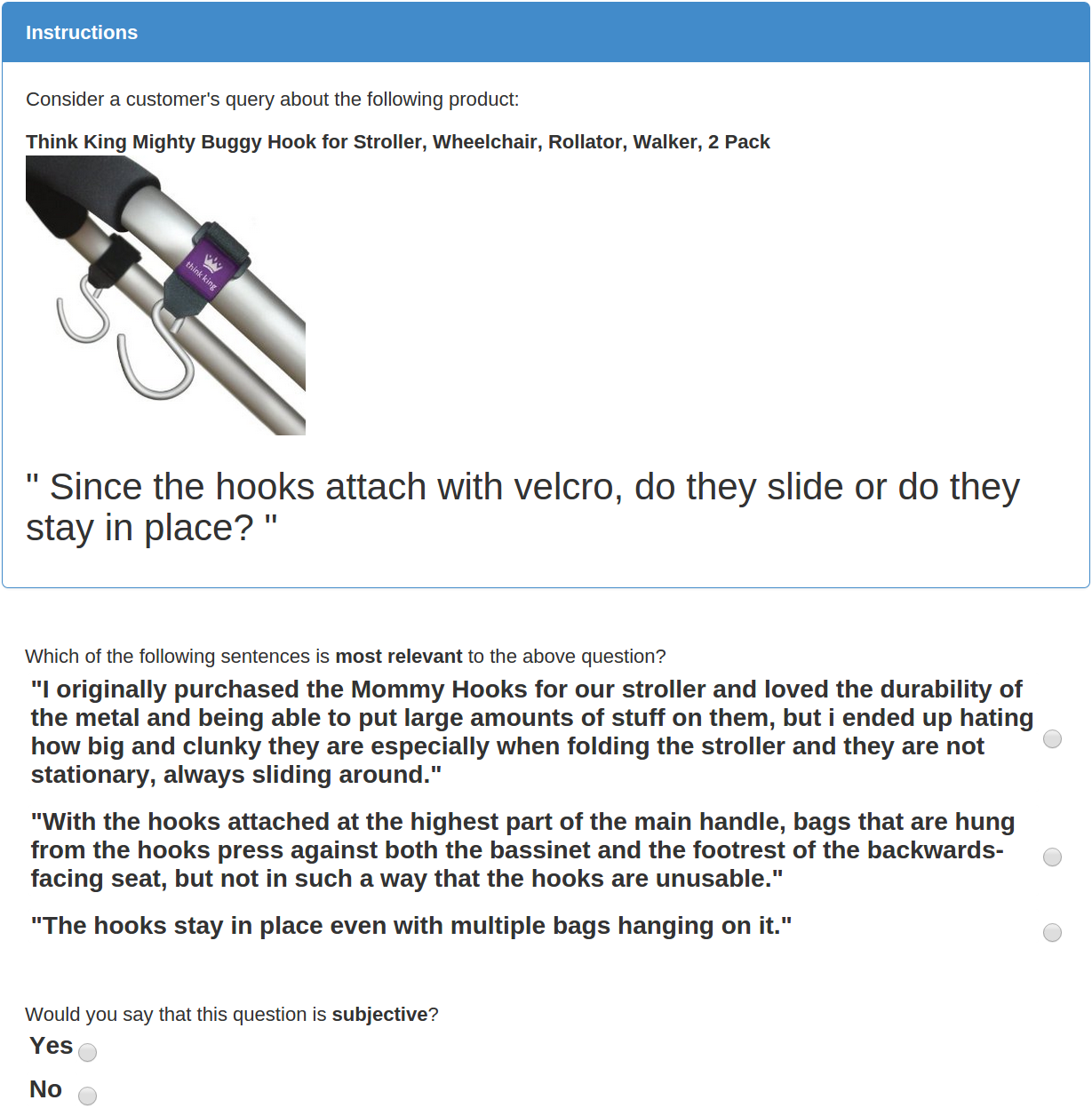}}
 \end{center}
\caption{A screenshot of our interface for user evaluation.\label{fig:interface}}
\end{figure}

We performed our evaluation using \emph{Amazon's Mechanical Turk}, using `master workers' to evaluate 100 queries from each of our five largest datasets, as well as one smaller dataset (\emph{baby}) to assess whether our method still performs well when less data is available for training. Workers were presented with a product's title, image, and a randomly selected query (binary or otherwise). We then presented them the top-ranked result from our method, as well as the top-ranked result using Okapi-BM25+/ROUGE measures (with tuned parameters, i.e., ro-L from Table \ref{tab:openended}); this represents a state-of-the-art `off-the-shelf' relevance ranking benchmark, with parameters tuned following best practices; it is also the most competitive baseline from Table \ref{tab:openended}. Results were shown to evaluators in a random order without labels, from which they had to select whichever they considered to be the most relevant.\footnote{We also showed a randomly selected result, and gave users the option to select \emph{no} result. We discarded cases with overlaps.} We also asked workers whether they considered a question to be `subjective' or not, in order to evaluate whether the subjectivity of the question impacts performance.
A screenshot of our interface is shown in Figure \ref{fig:interface}.

\begin{figure}[t]
\begin{center}
 \includegraphics{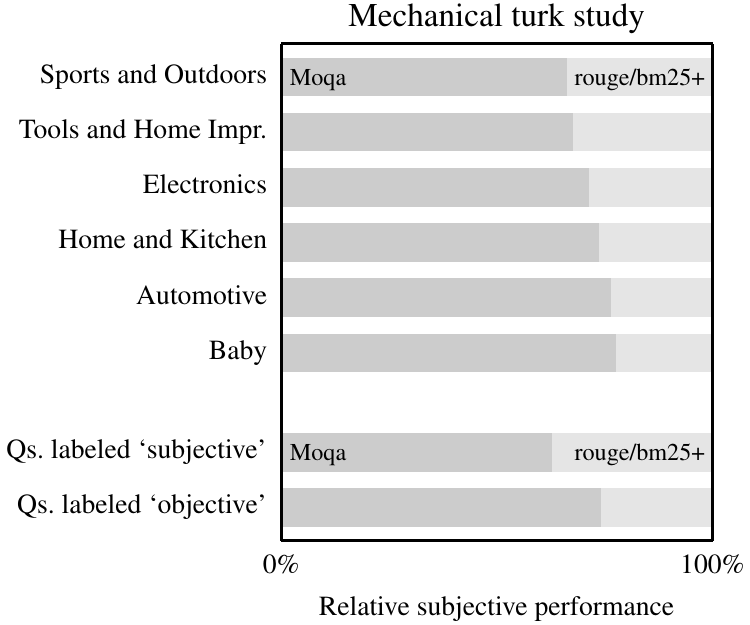}
 \end{center}
 \caption{User study. Bars indicate the fraction of times that opinions surfaced by \modelname{} are preferred over those of the strongest baseline (a tuned combination of BM25+ and the ROUGE score, ro-L from Section \ref{sec:baselines}). \label{fig:user_study}}
\end{figure}

Results of this evaluation are shown in Figure \ref{fig:user_study}. On average, \modelname{} was preferred in 73.1\% of instances across the six datasets we considered. This is a significant improvement; improvements were similar across datasets (between 66.2\% on Sports and Outdoors and 77.6\% on Baby), and for both subjective and objective queries (62.9\% vs.~74.1\%). Ultimately \modelname{} consistently outperforms our strongest baseline in terms of subjective performance, though relative performance seems to be about the same for objective and subjective queries, and across datasets.

\subsubsection{Examples}

\begin{figure*}[ht]
\begin{center}
\small
\begin{tabular}{p{0.82\linewidth}m{0.13\linewidth}}
\multicolumn{2}{c}{Binary model:}\\
\toprule
\parbox{\linewidth}{%
\textbf{Product:} Schwinn Searcher Bike (26-Inch, Silver) (\texttt{amazon.com/dp/B007CKH61C})\\
\textbf{Question:} ``Is this bike a medium? My daughter is 5'8''.''\\[-1mm]

\textbf{Ranked opinions and votes:} ``The seat was just a tad tall for my girl so we actually sawed a bit off of the seat pole so that it would sit a little lower.'' (yes, .698); ``The seat height and handlebars are easily adjustable.'' (yes, .771); ``This is a great bike for a tall person.'' (yes, .711)\\[-1mm]

\textbf{Response:} Yes (.722)\\
\textbf{Actual answer (labeled as `yes'):} My wife is 5'5'' and the seat is set pretty low, I think a female 5'8'' would fit well with the seat raised.
} & {\centering \includegraphics[width=\linewidth]{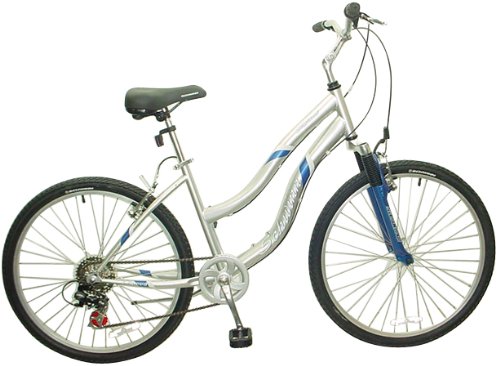}}\\
\midrule
\parbox{\linewidth}{%
\textbf{Product:}~Davis \& Sanford EXPLORERV Vista Explorer 60" Tripod (\texttt{amazon.com/dp/B000V7AF8E})\\
\textbf{Question:}~``Is this tripod better then the AmazonBasics 60-Inch Lightweight Tripod with Bag one?''\\[-1mm]
 
\textbf{Ranked opinions and votes:} ``However, if you are looking for a steady tripod, this product is not the product that you are looking for'' (no, .295); ``If you need a tripod for a camera or camcorder and are on a tight budget, this is the one for you.'' (yes, .901); ``This would probably work as a door stop at a gas station, but for any camera or spotting scope work I'd rather just lean over the hood of my pickup.'' (no, .463);\\[-1mm]

\textbf{Response:} Yes (.863)\\
\textbf{Actual answer (labeled as `yes'):} The 10 year warranty makes it much better and yes they do honor the warranty. I was sent a replacement when my failed.
} & \parbox{\linewidth}{\centering \includegraphics[width=0.75\linewidth]{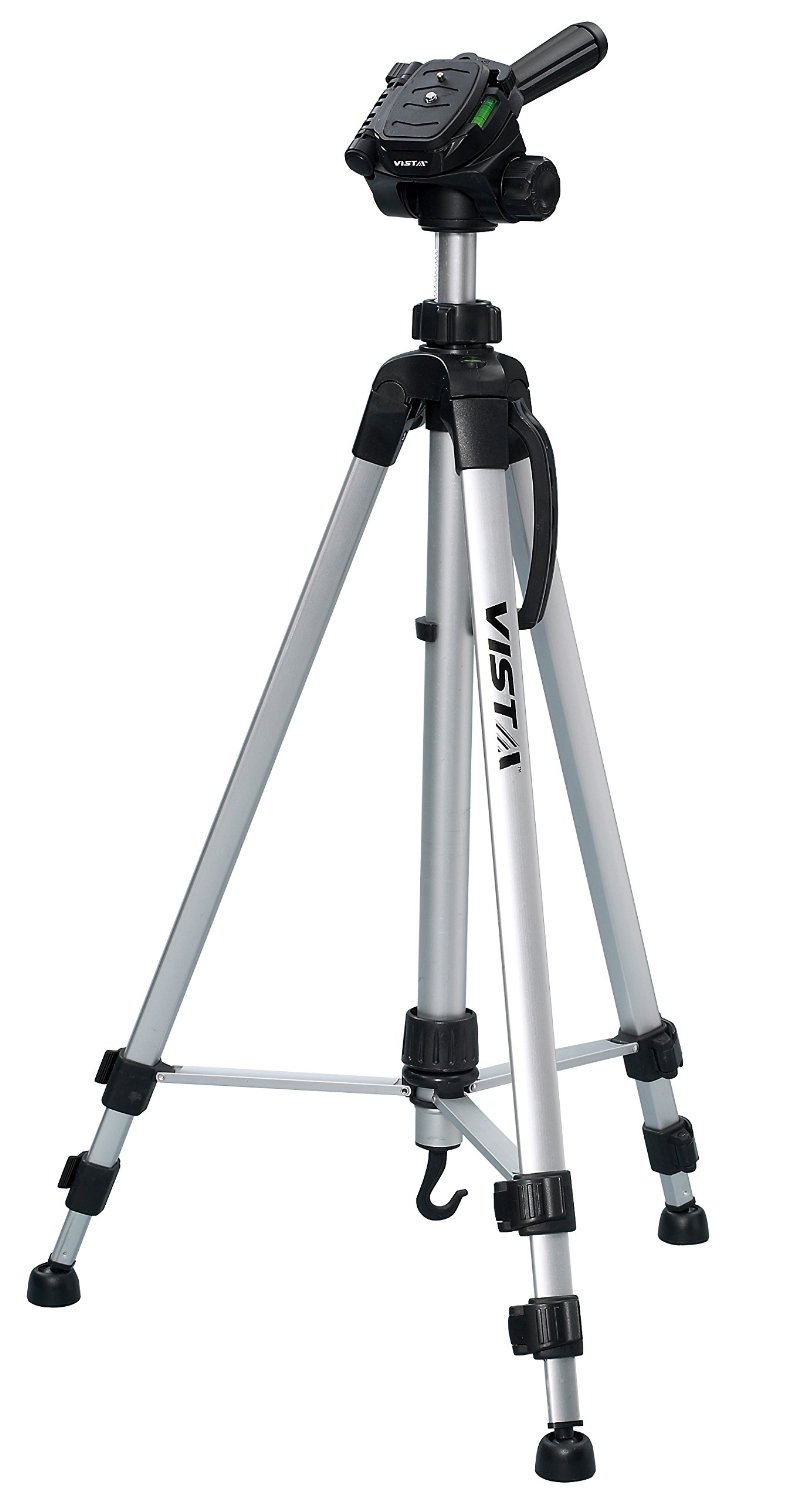}}\\
\bottomrule
\\
\multicolumn{2}{c}{Open-ended model:}\\
\toprule
\parbox{\linewidth}{%
\textbf{Product:}~Mommy's Helper Kid Keeper (\texttt{amazon.com/dp/B00081L2SU})\\
\textbf{Question:}~``I have a big two year old (30 lbs) who is very active and pretty strong. Will this harness fit him? Will there be any room to grow?''\\[-1mm]
 
\textbf{Ranked opinions:} ``So if you have big babies, this may not fit very long.''; ``They fit my boys okay for now, but I was really hoping they would fit around their torso for longer.''; ``I have a very active almost three year old who is huge.''\\[-1mm]

\textbf{Actual answer:} One of my two year olds is 36lbs and 36in tall. It fits him. I would like for there to be more room to grow, but it should fit for a while.
} & \parbox{\linewidth}{\centering \includegraphics[width=0.8\linewidth]{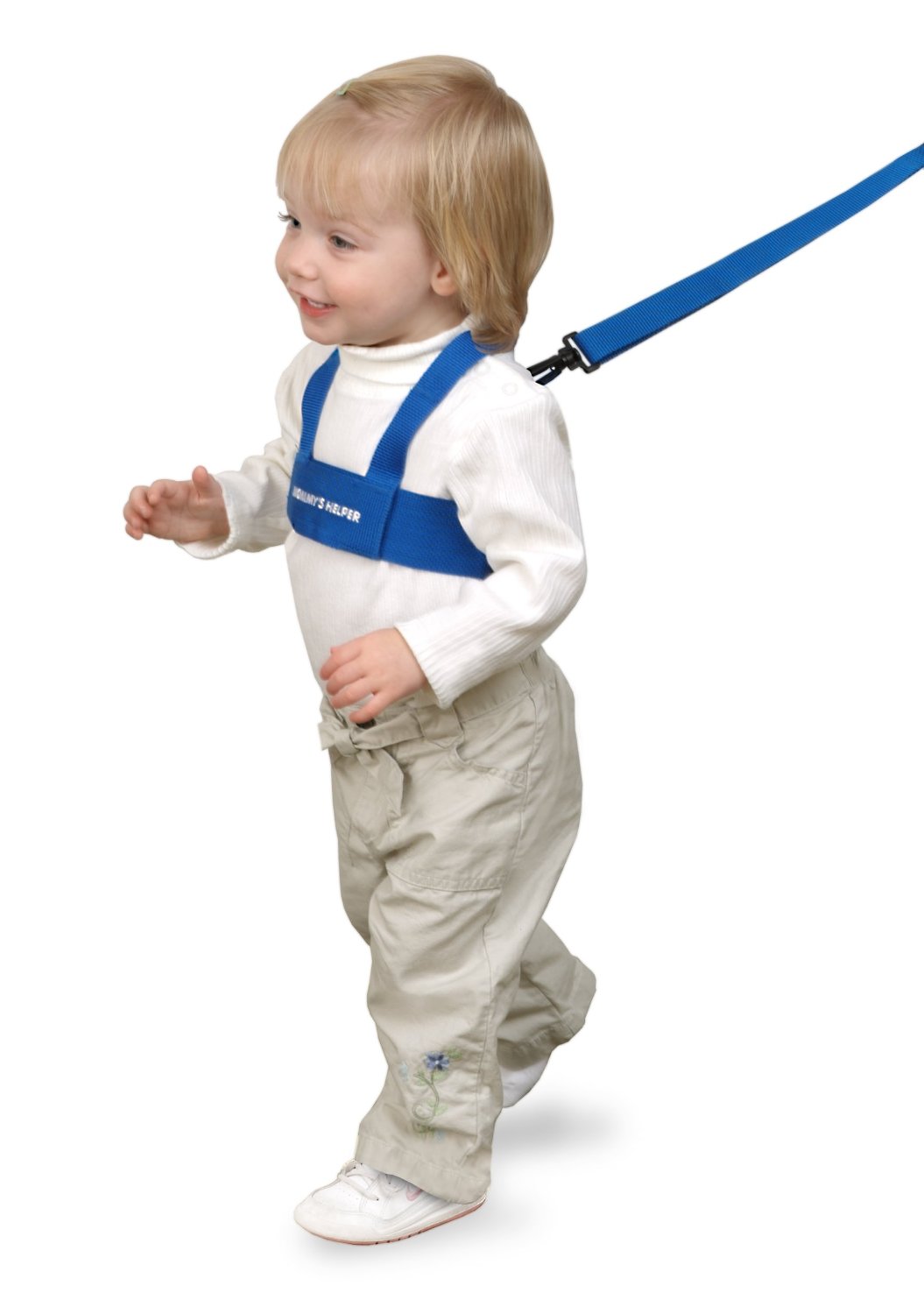}}\\
\midrule
\parbox{\linewidth}{%
\textbf{Product:}~Thermos 16 Oz Stainless Steel (\texttt{amazon.com/dp/B00FKPGEBO})\\
\textbf{Question:}~``how many hours does it keep hot and cold ?''\\[-1mm]
 
\textbf{Ranked opinions:} ``Does keep the coffee very hot for several hours.''; ``Keeps hot Beverages hot for a long time.''; ``I bought this to replace an aging one which was nearly identical to it on the outside, but which kept hot liquids hot for over 6 hours.''; ``Simple, sleek design, keeps the coffee hot for hours, and that's all I need.''; ``I tested it by placing boiling hot water in it and it did not keep it hot for 10 hrs.''; ``Overall, I found that it kept the water hot for about 3-4 hrs.''; \\[-1mm]

\textbf{Actual answer:} It doesn't, I returned the one I purchased.
} & \parbox{\linewidth}{\centering \includegraphics[width=0.5\linewidth]{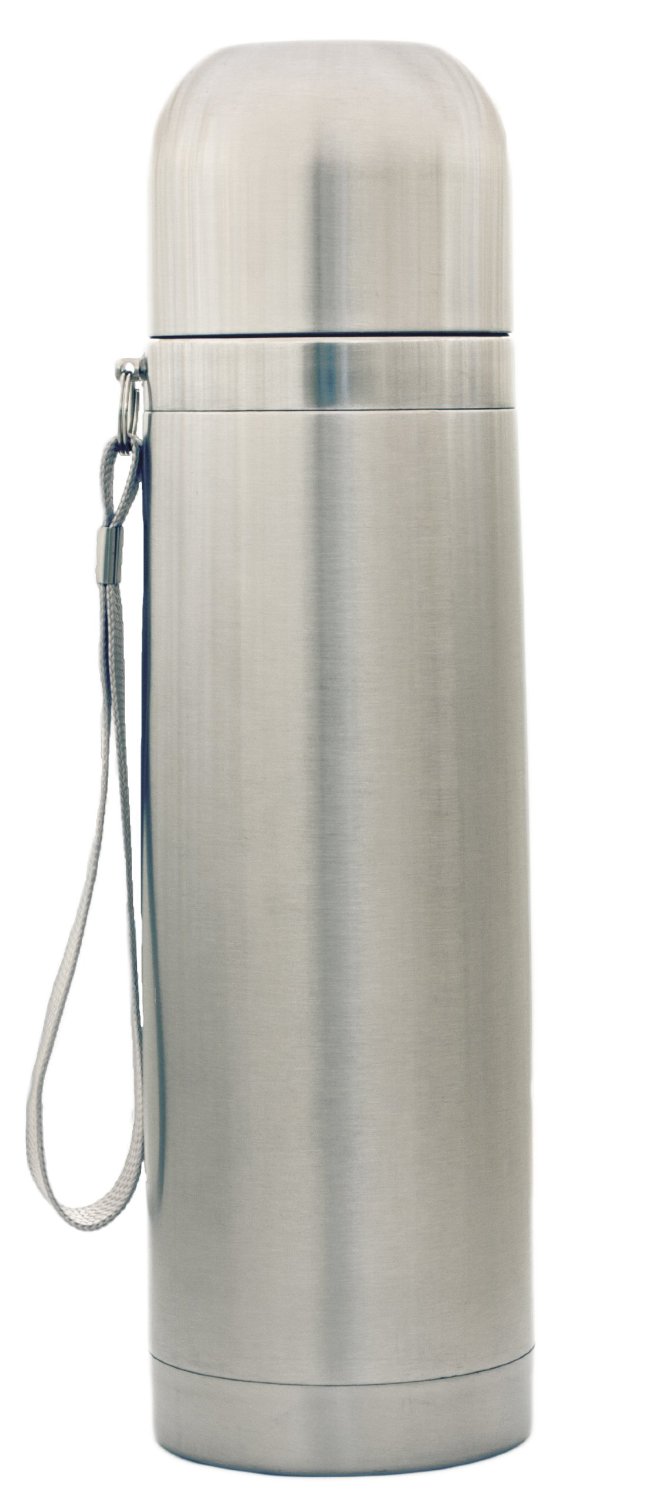}}\\
\bottomrule
\end{tabular}
\end{center}
\normalsize
\caption{Examples of opinions recommended by \modelname{}. The top two examples are generated by the binary model, the bottom two by the open-ended model. Note that none of these examples were available at training time, and only the question is provided as input (the true answer and its label are shown for comparison). Opinions are shown in decreasing order of relevance. Note in the second example that \emph{all} opinions get to vote in proportion to their relevance; in this case the many positive votes among less-relevant opinions outweigh the negative votes above, ultimately yielding a strong `yes' vote.\label{fig:examps}}
\end{figure*}

Finally, a few examples of the output produced by \modelname{} are shown in Figure \ref{fig:examps}. Note that none of these examples were available at training time, and only the question (along with the product being queried) are provided as input. These examples demonstrate a few features of \modelname{} and the data in question: First is the wide variety of products, questions, and opinions that are reflected in the data; this linguistic variability demonstrates the need for a model that \emph{learns} the notion of relevance from data. Second, the questions themselves (like the example from Figure \ref{fig:examp}) are quite different from those that could be answered through knowledge bases; even those that seem objective (e.g. ``how long does this stay hot?'') are met with a variety of responses representing different (and sometimes contradictory) experiences; thus reviews are the perfect source of data to capture this variety of views. Third is the heterogeneity between queries and opinions; words like ``girl'' and ``tall'' are identified as being relevant to ``daughter'' and ``medium,'' demonstrating the need for a flexible model that is capable of learning complicated semantics in general, and synonyms in particular.

Also note that while our bilinear model has many thousands of parameters, at test time relevance can be computed extremely efficiently, since in \eq{eq:approxtransform} we can project all reviews via $B$ in advance. Thus computing relevance takes only $O(K + |q| + |r|)$ (i.e., the number of projected dimensions plus the number of words in the query and review); in practice this allows us to answer queries in a few milliseconds, even for products with thousands of reviews.

\section{Discussion and Future Work}
\label{sec:discussion}

Surprisingly, performance for open-ended queries (Table \ref{tab:openended}) appears to be better than performance for binary queries (Table \ref{tab:yn}), both compared to random classification and to our strongest baseline, against our intuition that the latter task might be more difficult. There are a few reasons for this: One is simply that the task of differentiating the true answer from a (randomly selected) non-answer is `easier' than resolving a binary query; this explains why outperforming a random baseline is easier, but does not explain the higher relative improvement against baselines. For the latter, note that the main difference between our method and the strongest baseline is the use of a bilinear model; while a highly flexible model, it has far more parameters than baselines, meaning that a large dataset is required for training. Thus what we are seeing may simply be the benefit of having substantially more data available for training when considering open-ended questions.

Also surprising is that in our user study we obtained roughly equal performance on subjective vs.~objective queries. Partly this may be because subjective queries are simply `more difficult' to address, so that there is less separation between methods, though this would require a larger labeled dataset of subjective vs.~objective queries to evaluate quantitatively. In fact, contrary to expectation only around 20\% of queries were labeled as being `subjective' by workers. However the full story seems more complicated---queries such as ``how long does this stay hot?''~(Figure \ref{fig:examps}) are certainly labeled as being `objective' by human evaluators, though the variety of responses shows a more nuanced situation. Really, a large fraction of seemingly objective queries are met with contradictory answers representing different user experiences, which is exactly the class of questions that our method is designed to address.

\subsection{Future work}
We see several potential ways to extend \modelname{}.

First, while we have made extensive use of reviews, there is a wealth of additional information available on review websites that could potentially be used to address queries. One is rating information, which could improve performance on certain evaluative queries (though to an extent we already capture this information as our model is expressive enough to learn the polarity of sentiment words). Another is user information---the identity of the questioner and the reviewer could be used to learn better relevance models, both in terms of whether their opinions are aligned, or even to identify topical experts, as has been done with previous Q/A systems \cite{experts_qa,experts_qa2,jurczyk2007discovering,pal11b,anderson}.

In categories like electronics, a large fraction of queries are related to compatibility (e.g.~``will this product work with X?''). Addressing compatibility-related queries with user reviews is another promising avenue of future work---again, the massive number of potential product combinations means that large volumes of user reviews are potentially an ideal source of data to address such questions.
Although our system can already address such queries to some extent, ideally a model of compatibility-related queries would make use of additional information, for instance reviews of \emph{both} products being queried, or the fact that compatibility relationships tend to be symmetric, or even co-purchasing statistics as in \cite{McAPanLes15}.

Finally, since we are dealing with queries that are often subjective, we would like to handle the possibility that they may have multiple and potentially inconsistent answers. Currently we have selected the top-voted answer to each question as an `authoritative' response to be used at training time. But handling multiple, inconsistent answers could be valuable in several ways, for instance to automatically identify whether a question is subjective or contentious, or otherwise to generate relevance rankings that support a spectrum of subjective viewpoints.

\section{Conclusion}

We presented \modelname{}, a system that automatically responds to product-related queries by surfacing relevant consumer opinions. We achieved this by observing that a large corpus of previously-answered questions can be used to \emph{learn} the notion of relevance, in the sense that `relevant' opinions are those for which an accurate predictor can be trained to select the correct answer as a function of the question and the opinion. We cast this as a mixture-of-experts learning problem, where each opinion corresponds to an `expert' that gets to vote on the correct response, in proportion to its relevance. These relevance and voting functions are learned automatically and evaluated on a large training corpus of questions, answers, and reviews from \emph{Amazon}.

The main findings of our evaluation were as follows: First, reviews proved particularly effective as a source of data for answering product-related queries, outperforming other sources of text like product specifications; this demonstrates the value of \emph{personal experiences} in addressing users' queries. Second, we demonstrated the need to handle heterogeneity between various text sources (i.e., questions, reviews, and answers); our large corpus of training data allowed us to train a flexible bilinear model that it capable of automatically accounting for linguistic differences between text sources, outperforming hand-crafted word- and phrase-level relevance measures. Finally, we showed that \modelname{} is quantitatively able to address both binary and open-ended questions, and qualitatively that human evaluators prefer our learned notion of `relevance' over hand-crafted relevance measures.

\bibliographystyle{abbrv}

\end{document}